\documentclass[useAMS,usenatbib,usegraphicx]{mn2e}%doublespacing]{mn2e} %onecolumn
\makeatletter
\def\endfigure{\end@float}
\def\endfigure{\end@float}
\makeatother
\usepackage{afterpage}
\usepackage{float}
\usepackage{subfigure}

\usepackage{amsmath,amssymb}
\usepackage{mathpazo}
\usepackage{bm}
\usepackage{graphicx}
\usepackage{verbatim}
\usepackage{wrapfig}
\usepackage{makeidx}
\usepackage{mathrsfs}
\usepackage{amsfonts}
\usepackage{color}
\usepackage{comment} 
\def\Vec#1{\mbox{\boldmath $#1$}}
\begin{document}
\title[Detection of a filament connected to CL0016]{Detection of a filament connected to CL0016 with weak gravitational lensing}
\author[Y.Higuchi et al.]{Yuichi Higuchi$^{1}$\thanks{E-mail: yuichi.higuchi@nao.ac.jp}, Masamune Oguri$^{2, 3, 4}$, Masayuki Tanaka$^{5}$ and Junya Sakurai$^{5, 6}$\\
$^{1}$Astronomy Data Center, National Astronomical Observatory of Japan, 2-21-1 Osawa, Mitaka, Tokyo 181-8588, Japan\\
$^{2}$Research Center for the Early Universe, University of Tokyo, 7-3-1 Hongo, Bunkyo-ku, Tokyo 113-0033, Japan\\
$^{3}$Department of Physics, University of Tokyo, 7-3-1 Hongo, Bunkyo-ku, Tokyo 113-0033, Japan\\
$^{4}$Kavli Institute for the Physics and Mathematics of the Universe (Kavli IPMU, WPI), University of Tokyo, Chiba 277-8583, Japan\\
$^{5}$Optical and Infrared Astronomy Division, National Astronomical Observatory of Japan, 2-21-1 Osawa, Mitaka, Tokyo 181-8588, Japan\\
$^{6}$Department of Astronomical Science, SOKENDAI(The Graduate University for Advanced Studies), Mitaka, Tokyo, 181-8588, Japan\\
}
\maketitle

\begin{abstract}
We report on the weak lensing detection of a filament between two galaxy clusters at $z=0.55$, CL0015.9+1609 and RX J0018.3+1618. We conduct weak lensing analysis of deep multi-band Subaru/Suprime-Cam images with $Lensfit$. The weak lensing signals from the filament are contaminated by signals from the adjacent massive clusters and we statistically subtract the cluster component using two different methods. Both methods yield consistent shear profiles on the filament with $\ga2\sigma$ significance and the average surface mass density of the filament is $\langle\Sigma\rangle=(3.20\pm0.10)\times10^{14}h$ M$_\odot$Mpc$^{-2}$, which is in broad agreement with previous studies. On-going surveys such as Hyper Suprime-Cam will identify more filaments, which will serve as a new probe of structure formation in the Universe.
\end{abstract}

%%%look up list of allowable keywords for this section

%%%%%%%%%%%%%%%%%%%%%%%%introduction %%%%%%%%%%%%%%%%%%%%%%%%
\section{Introduction}
Recent observations have revealed that the Universe is dominated by unknown components called dark matter and dark energy \citep{2013ApJS..208...19H, 2014ApJ...795...44R, 2015arXiv150201589P}.  The theoretical model called $\Lambda$ Cold Dark Matter ($\Lambda$CDM) model, which contains these components, has been confronted with several observational tests such as correlation functions (e.g., \citealt{2005MNRAS.363.1329B, 2006PhRvD..74l3507T, 2006A&A...459..375H, 2013MNRAS.432.2654M}) and Minkowski functionals (e.g., \citealt{2008MNRAS.389.1439H, 2013PhRvD..88l3002P}). These statistical observables are in excellent agreement with theoretical predictions from the $\Lambda$CDM model. However, these tests only use ensemble average properties of the matter distribution. Complementary tests can be done with voids and filaments, which are important constitutes of the large-scale structure of the Universe. Voids are underdense regions whose properties have been studied theoretically (e.g., \citealt{2006MNRAS.366..467F, 2009JCAP...01..010K, 2013MNRAS.432.1021H, 2013ApJ...762L..20K, 2015MNRAS.446L...1S}). On the other hand, filaments  are slightly overdense regions connecting massive haloes. 

Large $N$-body simulations indicate that filaments are formed in the process of merging and accretion \citep{1996Natur.380..603B,2005MNRAS.364.1105S,2005MNRAS.359..272C}. Therefore, filaments affect properties of nearly haloes. The properties of haloes residing with filaments are well studied through $N$-body simulations \citep{2012MNRAS.427.3320C,2014MNRAS.440L..46A,2014MNRAS.441..745H}. In order to find such filamentary structures, a variety of algorithms have been developed \citep{2002SPIE.4847...86M, 2003MNRAS.345..529S, 2006astro.ph..2628S, 2010MNRAS.401.2257M}.

Although we can see cosmic web in simulations, filamentary structures are hard to be detected observationally due to their low overdensity \citep{2008MNRAS.383.1655S, 2008MNRAS.385.1431H, 2014arXiv1402.3302C}. Some studies claimed the detection of filaments with X-ray \citep{1995A&A...302L...9B, 2001ApJ...563..673T, 2002ApJ...572L.127F, 2003A&A...403L..29D, 2008PhDT.......221W}, although it is difficult to distinguish whether those signals come from filaments or from haloes at the edge of the filaments. 

While X-ray probes only hot intergalactic medium, weak gravitational lensing can probe not only baryon but also dark matter which occupies a dominant part of a filament mass. However, the detection of filaments with weak lensing have not been very successful. While weak lensing detections of filaments between clusters have been claimed (e.g., \citealt{1998ApJ...497L..61C, 1998astro.ph..9268K,2002ApJ...568..141G,2004ogci.conf...34D}), it has also been suggested that these detection signals might have come from systematics of the analysis \citep{2004A&A...422..407G, 2008MNRAS.385.1431H}. Recently more secure detections of filaments with weak gravitational lensing have been reported in the systems Abell 222/223  \citep{2012Natur.487..202D} and MACSJ0717.5+3745 \citep{2012MNRAS.426.3369J}. However, a lager sample is needed in order to test the $\Lambda$CDM model with large-scale structures, and also to investigate observational properties of filaments. In this paper, we report the new detection of a filament with weak gravitational lensing.

This paper is organized as follows. In Section~\ref{sec.analysis},  we describe the lensing analysis methods with a focus on the basics of weak lensing and halo properties. In Section~\ref{sec.ana.obs}, we describe our data analysis method, including the selection of background galaxies, the lensing analysis with $Lensfit$, the reconstruction of the filament mass distribution and the evaluation of contaminations from clusters. In Section~\ref{sec.result}, we show the results of lensing analysis. We summarize our result in Section~\ref{sec.conclusion}.

 In this paper, cosmological parameters are set according to the result of the Planck satellite \citep{2013arXiv1303.5076P}; Hubble constant H$_0=67.3$ km/s/Mpc, cosmological constant $\Omega_{\rm m}=0.315$, dark energy $\Omega_\Lambda=0.685$.

%%%%%%%%%%%%%%%%%%%%%%filament detection%%%%%%%%%%%%%%%%%%%%%%%
\section{Theories for weak lensing}
\label{sec.analysis}
First, we summarize the theory of weak gravitational lensing. We follow the formalism outlined in \citet{2001PhR...340..291B}.
%%%%%estimate lensing signal from ellipticity
\subsection{Estimation of lensing signals from images}
Weak lensing signals are obtained from observed galaxy ellipticities. However, convergence $\kappa(\theta)$ and shear $\gamma(\theta)$ are not directly related to galaxy ellipticities. The galaxy ellipticity of a lensed object depends on the reduced shear ${\rm g}$ defined as   
\begin{equation}
{\rm g}(\Vec{\theta})=\frac{\gamma(\Vec{\theta})}{1-\kappa(\Vec{\theta})}.
\end{equation}
In the weak lensing limit, the reduced shear is related to the galaxy ellipticity $\varepsilon$ as
\begin{equation}
\varepsilon \sim \varepsilon^{\rm s}+{\rm g},  
\end{equation}
where $\varepsilon^{\rm s}$ is an intrinsic ellipticity of the galaxy. On the basis of the cosmological principle, we can assume that orientations of galaxies are random, i.e., the ensemble average of intrinsic ellipticities should be zero. Therefore, the average observed ellipticity is related to reduced shear as
 \begin{equation}
\langle\varepsilon\rangle=\langle{\rm g}\rangle \sim\langle\gamma\rangle,
\end{equation} 
 where $\langle\cdots\rangle$ indicates ensemble average in a bin. 
 
We need to calculate the lensing depth of source galaxies in order to estimate physical quantities such as cluster masses from weak lensing signals. The lensing amplitude is proportional to the distance ratio averaged over the population of source galaxies, 
\begin{equation}
\left\langle\frac{\rm D_{\rm ls}}{\rm D_{\rm s}}\right\rangle=\int {\rm d}z\frac{{\rm dp}}{{\rm d}z}\frac{\rm D_{\rm ls}}{\rm D_{\rm s}},
\label{eq.lensamp}
\end{equation}
where ${\rm D_{\rm ls}}$ and ${\rm D_{\rm s}}$ are the angular diameter distances from the lens to the source and from the observer to the source, respectively. ${\rm dp}/{\rm d}z$ is the probability distribution function of redshifts of source galaxies. The critical projected mass density is described with the lensing amplitude as
\begin{equation}
\Sigma_{\rm cr}=\frac{\rm c^2}{4\pi {\rm G}}{\rm D_l^{-1}}\left\langle\frac{\rm D_{ls}}{\rm D_s}\right\rangle^{-1},
\end{equation}
where c is speed of light and G is gravitational constant.
In this paper, we assume that all galaxies are located at this distance.

%%%%%%%%%%%%%%%%%%%%%%fitting%%%%%%%%%%%%%%%%%%%%%%%%%%%%%%
\subsection{Model of filament profile}
\label{sec.filpro}
The filament profile has been studied through both simulations and observations \citep{2005MNRAS.359..272C, 2012Natur.487..202D, 2014MNRAS.441..745H}.
These studies showed that the convergence profile of filaments is described as a function of distance $\theta$ from the line which connects between haloes on the sky plane (hereafter halo-halo axis) as \citep{2005MNRAS.359..272C}
\begin{equation}
\kappa(\theta)=\frac{\kappa_0}{1+\left(\theta/\theta_c\right)^2},
\label{eq.filamentmodel}
\end{equation}
where $\kappa_0$ describes the convergence value at $\theta=0$ and $\theta_c$ is the scale length of a filament. 
For spherically symmetric objects, it is common to use reduced shear values defined relative to the centre of objects as
\begin{equation}
\left(
\begin{array}{cc}
{\rm g}_+   \\
{\rm g}_\times     
\end{array}
\right)
=
\left(
\begin{array}{cc}
-\mathrm{cos}2\phi&-\mathrm{sin}2\phi \\
-\mathrm{sin}2\phi& \mathrm{cos}2\phi \\
\end{array}
\right)
\left(
\begin{array}{c}
{\rm g}_1\\
{\rm g}_2   
\end{array}
\right),
\label{angle}
\end{equation}
where $\phi$ is the angle between axis-$\theta_1$ and $\Vec{\theta}$. ${\rm g}_+$ and ${\rm g}_\times$ are tangential and cross reduced shear components, respectively. However, filaments are more like axial symmetric. Therefore following \citet{2014MNRAS.441..745H}, for filaments we define the "tangential shear" at each point as ${\rm g}_+$ relative to the closest point on the halo-halo axis to the point. In this definition, the tangential shear profile of filaments takes negative values near the halo-halo axis and positive values at large radii (see Section~\ref{subsec.estclus}).

%%%%%%%%%%%%halo %%%%%%%%%%%%%%%%%%%%%%%
%describe model halo profile (NFW parameters) based on oguri et al 
\subsection{NFW profile}
Weak lensing signals from filaments are contaminated by signals from nearby clusters. We estimate the effect from clusters by using the cluster mass profile, referred as the Navarro, Frenk and White (hereafter NFW) model \citep{1997ApJ...490..493N}, which has well been studied in simulations and observations \citep{2008MNRAS.387..536G, 2010PASJ...62..811O, 2012MNRAS.420.3213O}. These studies indicate that the three-dimensional mass profile of clusters declines as $r^{-1}$ inside the scale $r_s$ and de- clines as $r^{-3}$ outside $r_s$. The NFW profile is described as
\begin{equation}
\rho(\mathrm{r})= \left\{ 
\begin{array}{ll}
\vspace{0.2cm}\frac{\rho_s}{r/r_s(1+r/r_s)^2} & (r<r_{\mathrm{vir}})\\
\vspace{0.2cm}0& otherwise, 
\end{array}
\right.
\label{massdensity1}
\end{equation}
where $r_s$ and $r_\mathrm{vir}$ are the scale radius and virial radius, respectively. The characteristic density $\rho_s$ is defined as
\begin{equation}
\rho_s=\frac{\Delta_\mathrm{vir}(z)\bar{\rho}(z){\rm c}_\mathrm{vir}^3}{3\left[\mathrm{ln}\left(1+{\rm c}_\mathrm{vir}\right)-{\rm c}_\mathrm{vir}/\left(1+{\rm c}_\mathrm{vir}\right)\right]},
\end{equation}
where $\Delta_\mathrm{vir}(z)$ is the non-linear overdensity predicted by the spherical collapse model.
We parametrize the mass profile with two parameters. One is the concentration parameter:
\begin{equation}
{\rm c}_\mathrm{vir}=\frac{r_\mathrm{vir}}{r_s},
\end{equation}
and the other is the virial mass:
 \begin{equation}
\mathrm{M}_\mathrm{vir}=\frac{4\pi}{3}\Delta_\mathrm{vir}\left(z\right)\bar{\rho}(z)r_\mathrm{vir}^3.
\label{eq.mvir2}
\end{equation}

The analytical lensing profile of the NFW profile is well studied (see \citealt{2000ApJ...534...34W}). In order to constrain the cluster density profiles, we fit the lensing signals with the two fitting parameters ${\rm c}_\mathrm{vir}$ and $\mathrm{M}_\mathrm{vir}$ by minimizing a chi-square value defined as
\begin{equation}
\chi^2=\sum_i \frac{\left\{\mathrm{g}_{+, \mathrm{im}}(\theta_i)-\mathrm{g}_{+, \mathrm{ana}}(\theta_i)\right\}^2}{\sigma_{\mathrm{g}_{+, \mathrm{im}}}^2},
\label{eq.chisquare}
\end{equation}
where $\mathrm{g}_{+, \mathrm{im}}(\theta_i)$ and $\mathrm{g}_{+, \mathrm{ana}}(\theta_i)$ are circular-averaged tangential shear values at $i$-th bin estimated from data and model, respectively, and $\sigma_{\mathrm{g}_{+, \mathrm{im}}}$ is dispersion in $i$-th bin estimated from data.

%%%%%%%%%%%%%%%%% data analysis %%%%%%%%%%%%%%%%%%%%%%%%%%%%%%%%%%%%
\section{Data and analysis method}
\label{sec.ana.obs}
In this section, we summarize the data, weak lensing analysis methods and the methods for subtracting the cluster components.

\subsection{Data reduction}

\begin{figure}
\begin{center}
\includegraphics[width=8cm]{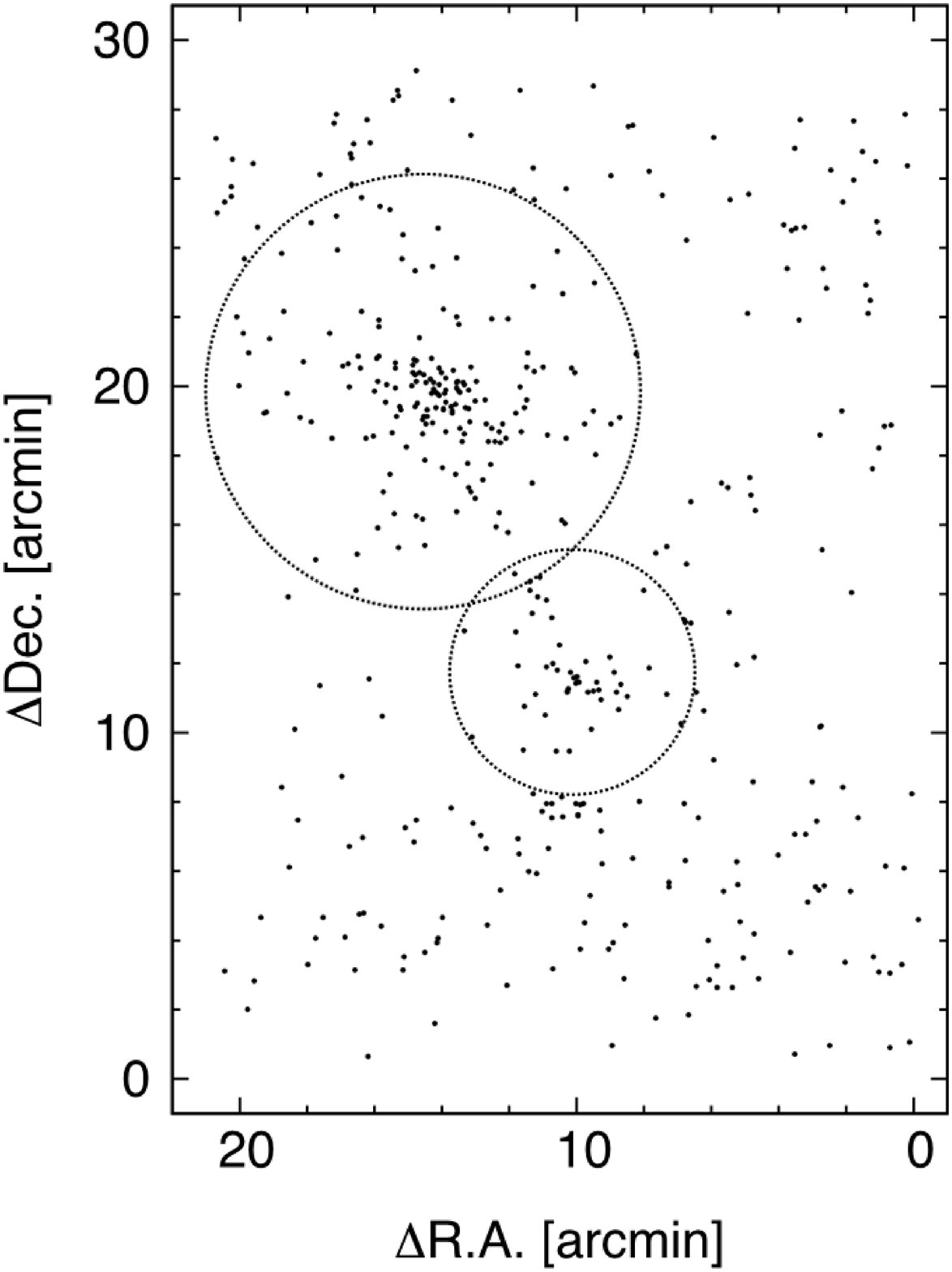}
\caption{Galaxy distributions at the cluster redshift. The filled points show galaxies in the photometric redshift range between $0.5<z_{\rm photo}-\sigma$ and $ z_{\rm photo}+\sigma<0.6$ where $\sigma$ is the photo-z estimation error. The circles show the viral radii of two clusters estimated from profile fitting in Section~\ref{sec.ana.obs}.}
\label{fig.gal_dis}
\end{center}
\end{figure}

We analyze the mass distribution around CL0015.9+1609 (hereafter CL0016) and RX J0018.3+1618 (hereafter RX J0018) at redshift $z=0.55$  \citep{1996ApJ...473L..67C, 2005PASJ...57..309K, 2005MNRAS.362..268T, 2009A&A...505L...9T}. These previous studies show that the galaxy distribution between these two clusters exhibit filamentary structure (see Figure~\ref{fig.gal_dis}). In the paper, we use the data taken with the Subaru Prime Focus Camera (Suprime-Cam; \citealt{2002PASJ...54..833M}) mounted on the Subaru Telescope during the nights of 2003 September 25-26 \citep{2005PASJ...57..309K, 2005MNRAS.362..268T}. We retrieve these data from the data archive system, SMOKA \citep{2002ASPC..281..298B}.  The field of view of the Suprime-Cam is $34\times27$ arcmin$^2$ with the pixel scale of $0.20$ arcsec. It is observed through several broad-band filters ($BVRi'z'$) down to limiting magnitude $B=26.9$, $V=26.2$, $R=26.0$, $i'=25.9$ and $z'=24.6$, respectively. The seeing size is about $0".65$ (FWHM). For more details, see \citet{2005PASJ...57..309K} and \citet{2005MNRAS.362..268T}. 

The data are reduced with a modified version of the LSST stack, following the standard procedures of bias subtraction, flat-fielding, distortion correction, sky subtraction and coadding to generate the stacked image from the mosaic images. The flat frames are constructed with the dithered science frames. The MAG\_AUTO in SExtractor is used as a measure of the total magnitude \citep{1996A&AS..117..393B}.

For weak lensing analysis, a precise PSF model is crucial and we select PSF reference stars as follows.
Objects are detected in the stacked $i'$-band image by using SExtractor. We use the same detection criteria used in \citet{2005PASJ...57..309K}. Stars are then selected in a standard way by identifying the stellar sequence in magnitude m$_{i'}$ vs half light radius r$_\mathrm{h}$ plane. To be specific, we apply $20.5\leq \mathrm{m}_i'\leq23.5$, r$_\mathrm{h}\leq0.75$ arcsec and more than 20$\sigma$ above the median sky. In addition, STAR\_CLASS$\geq$0.95 and FLAG$=0$ in SExtractor are also imposed. We set the fifth oder polynomial coefficients to fit the spatial variation of PSFs. For calculating the astrometric shifts between different shots, stars with 20$\sigma$ are used.

Since contaminations by galaxies near the cluster redshift dilute weak lensing signals, for weak lensing analysis we select only background galaxies using photometric redshifts. We use the photometric redshift catalog constructed in \citet{2005MNRAS.362..268T} with the photometric redshift code of \citet{1999MNRAS.302..152K}. Specifically, galaxies in photometric redshift range $z_{\rm photo}-\sigma\geq0.6$ are selected as background galaxies based on the photometric redshift catalogue, where $\sigma$ is an error for each photometric redshift estimation. The photometric redshifts of galaxies are compared with spectroscopical data \citep{1995ApJ...448L..93H, 1997ApJS..109...45M, 1998ApJ...497..645H}. The mean difference between spectroscopic redshift $z_\mathrm{spec}$ and photometric redshift is $z_\mathrm{photo}-z_\mathrm{spec}\sim0.004$. Our photometric redshifts tend to be underestimated for galaxies with $V-i'\sim1.5$  about $\Delta z\sim-0.1$ \citep{1999MNRAS.302..152K}. In order to investigate the effect of this tendency, we estimate lensing signals with several different redshift selection criteria. However, we do not find any significant change of our result.

\begin{figure}
\begin{center}
\includegraphics[width=9cm]{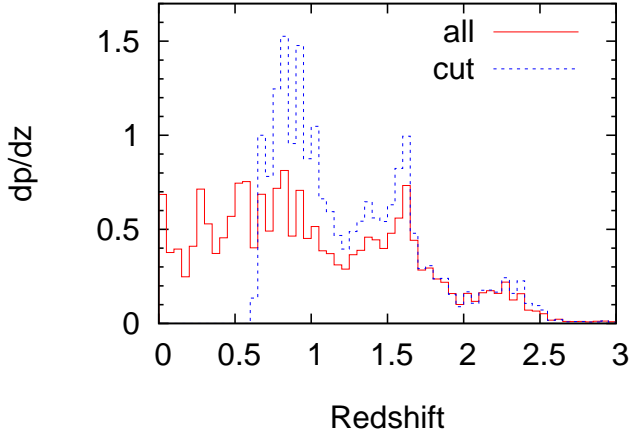}
\caption{Photometric redshift distributions of galaxies. Solid and dotted lines show the distribution for all galaxies and background galaxies defined by $z_{\rm photo}-\sigma\geq0.6$, respectively.}
\label{fig.galdishist}
\end{center}
\end{figure}

We use the code {\it Lensfit} \citep{2007MNRAS.382..315M, 2008MNRAS.390..149K, 2013MNRAS.429.2858M} for estimating the lensing signals. Lensing signals are estimated with the $i'$-band mosaic data. We choose $40\times40$ pixel$^2$ as a postage stamp size in order to estimate galaxy ellipticities with $Lensfit$. To eliminate galaxies that are not suitable for measuring lensing signals, e.g., blended galaxies, lensing signals are estimated with only galaxies with the flag $fitclass=0$ in $Lensfit$. The total number of galaxies used for weak lensing analysis is $28,315$. For each galaxy, we assign a weight defined by (e.g., \citealt{2013MNRAS.429.2858M})
\begin{equation}
w_{\rm g}=\left[\frac{\sigma_\varepsilon^2\epsilon_\mathrm{max}^2}{\varepsilon_\mathrm{max}^2-2\sigma_\varepsilon^2}+\sigma_\mathrm{pop}\right]^{-1},
\label{eq.weightgal}
\end{equation}
where $\sigma_\varepsilon$ is the one-dimensional variance in ellipticity of the likelihood surface, $\sigma_\mathrm{pop}$ is the one-dimensional variance of the distribution of ellipticity of the galaxy samples, and $\varepsilon_{\rm max}$ is the maximum allowed ellipticity. The reduced shear in the $n$-th bin is calculated with the weight as \citep{1995A&A...297..287S, 2001PhR...340..291B}
\begin{equation}
\langle{\rm g}_\mu\rangle\left(\theta_n\right)=\frac{\sum^{N_g}_{i=1} w_{{\rm g}, i} \varepsilon_{\mu, i}}{\sum^{N_g}_{i=1} w_{{\rm g}, i}},
\end{equation}
where $\mu$ indicates tangential or cross shear components. The convergence value at each point is obtained with galaxy ellipticity as
\begin{equation}
\kappa(\Vec{\theta})=\frac{1}{{\rm n_g}\pi}\sum_{i}{\rm Re}\left[\tilde{\mathscr{D}}^*\left(\Vec{\theta}-\Vec{\theta}_i\right)w_{{\rm g}, i}\varepsilon_i\right]/\sum_i w_{{\rm g}, i},
\label{eq.conv2}
\end{equation}
where  ${\rm n_g}$ is a number density of galaxies, and $\tilde{\mathscr{D}}$ is the kernel defined as
\begin{equation}
\tilde{\mathscr{D}}\left(\Vec{\theta}\right)=\left[1-\left(1+\frac{|\Vec{\theta}|^2}{\theta_{\rm s}^2}\right){\rm exp}\left(-\frac{|\Vec{\theta}|^2}{\theta_{\rm s}^2}\right)\right]\mathscr{D}(\Vec{\theta}),
\label{eq.kernel1}
\end{equation}
where $\theta_{\rm s}$ is a smoothing scale. The statistical error in the shear measurement for each bin is computed from the weighted average of the variance of the shear $\sigma_{\rm g}^2$ (see \citealt{2010PASJ...62..811O, 2012MNRAS.420.3213O}). For this purpose, we identify 20 neighboring galaxies in the magnitude-galaxy scale length plane. The variance, which is defined as $\sigma_{\rm g}^2=\sigma_{\rm g_1}^2+\sigma_{\rm g_2}^2$, is computed over the neighboring samples. Using the statistical error of shear for the $i$-th galaxy $\sigma_{{\rm g}, i}$, the uncertainty on the tangential shear in each bin is estimated with the weight as
\begin{equation}
\sigma_{{\rm g}_+}^2(\theta_n)=\frac{1}{2}\frac{\sum_i w_{{\rm g},i}^2\sigma_{{\rm g},i}^2}{\left(\sum_i w_{{\rm g}, i}\right)^2}.
\label{eq.variance}
\end{equation}

The factor 1/2 accounts for the fact that $\sigma_{{\rm g},i}$ is the rms for sum of two distortion components. We assume that the correlation between different bins is negligibly small, i.e., the main source of the statistical error comes from the intrinsic ellipticities. 

In order to quantify the significance of the filament detection, we estimate the difference of the profiles from a null profile, i.e., $\mathrm{g}_{+, \mathrm{ana}}(\theta_i)=0$ in equation~(\ref{eq.chisquare}). We estimate the significance by comparing the estimated chi-square values with the chi-square distribution. In this paper, we use filament shear profiles up to $15$ arcmin.

In the shear measurement, we select galaxies in the photometric redshift range $z-\sigma\geq0.6$ and estimate the lensing depth for the galaxies with equation~(\ref{eq.lensamp}). The value of the lensing depth $\left\langle{\rm D_{ls}}/{\rm D_s}\right\rangle$ is 0.419. Figure~\ref{fig.galdishist} shows the redshift distribution of galaxies. 

%%%%back ground estimation
\subsection{Estimation of the effect of the clusters}
\label{subsec.estclus}

\begin{figure}
\begin{center}
\includegraphics[width=9cm]{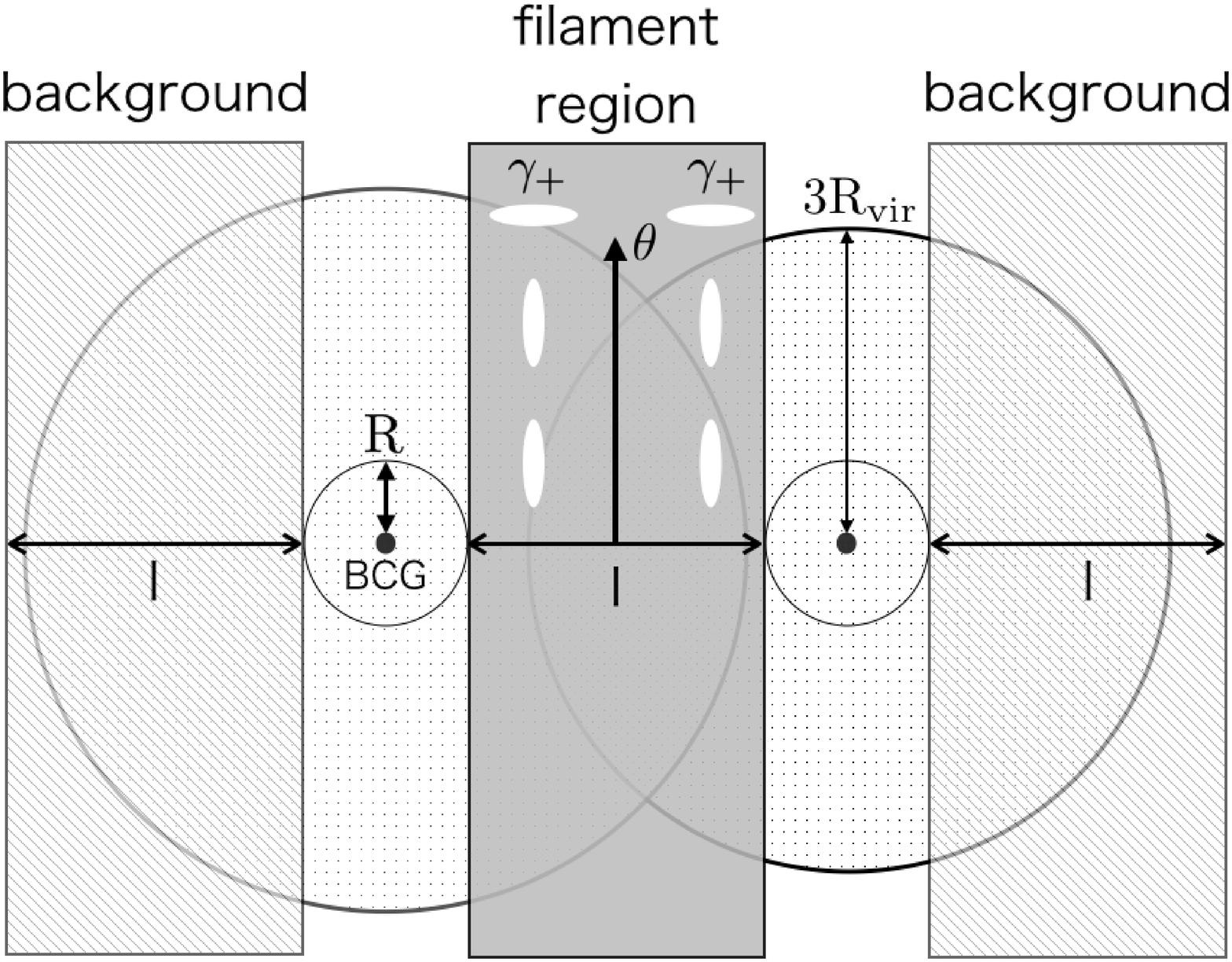}
\caption{Definition of the regions and explanation for the background estimation. The black points show the locations of BCGs. A region between clusters and outside ${\rm R}$ from each BCG are defined as the filament region (shaded region). In the first method for a background estimation, shear profiles inside $3{\rm R_{vir}}$ from BCGs (dotted regions) are estimated with the analytical models and subtracted. In the second method, the regions on the side opposite to the filament, where regions inside ${\rm R}$ from the each BCG are excluded, are defined as background (striped regions). Ellipses in the filament region indicate the distorted directions of images of background galaxies by a mass distribution of the filament.}
\label{fig.exreg}
\end{center}
\end{figure}

\begin{figure*}
\begin{center}
\subfigure{\includegraphics[width=0.8\columnwidth]{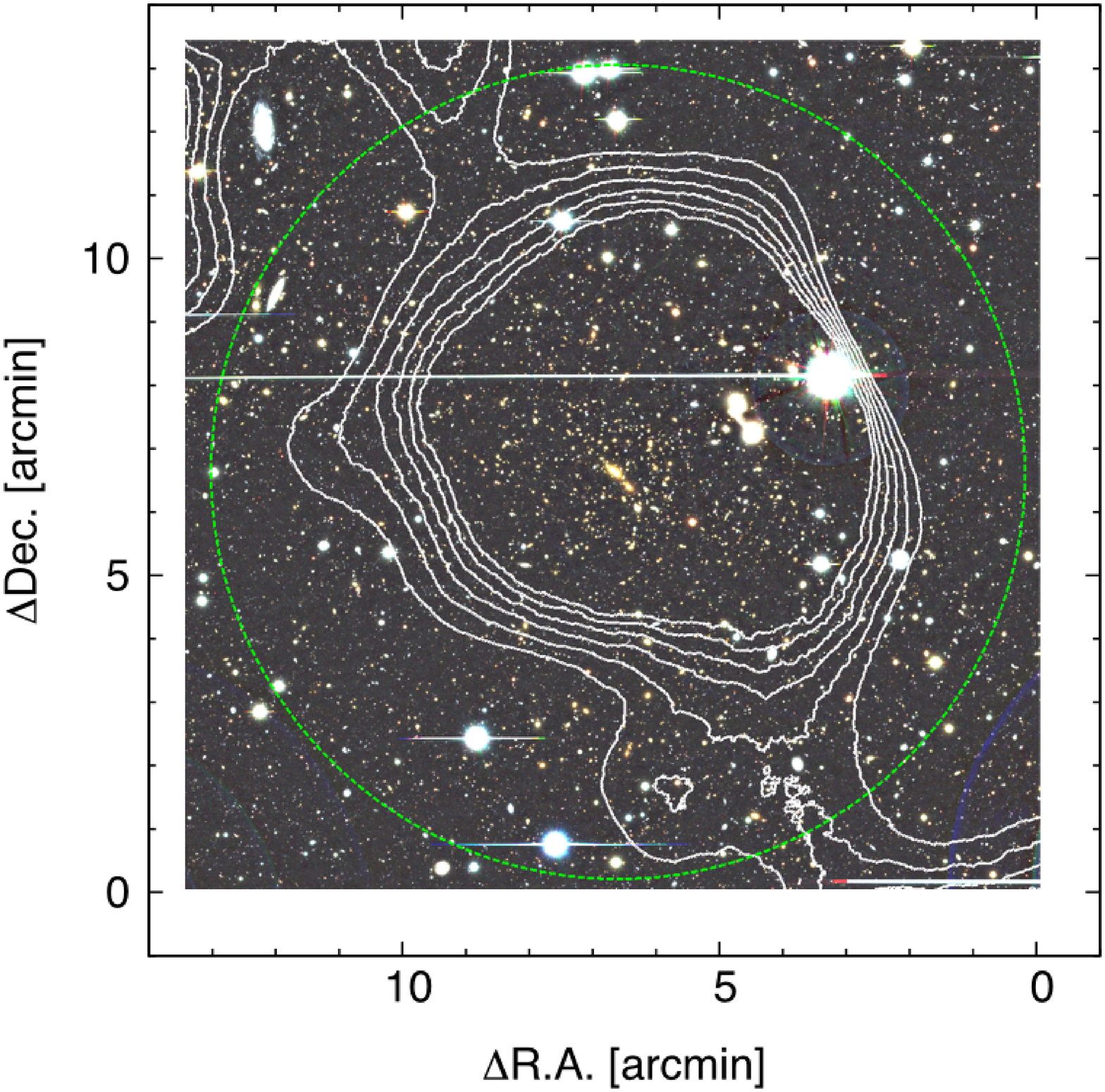}\label{fig.cl0016_contour}}\hspace{1cm}
\subfigure{\includegraphics[width=1.1\columnwidth]{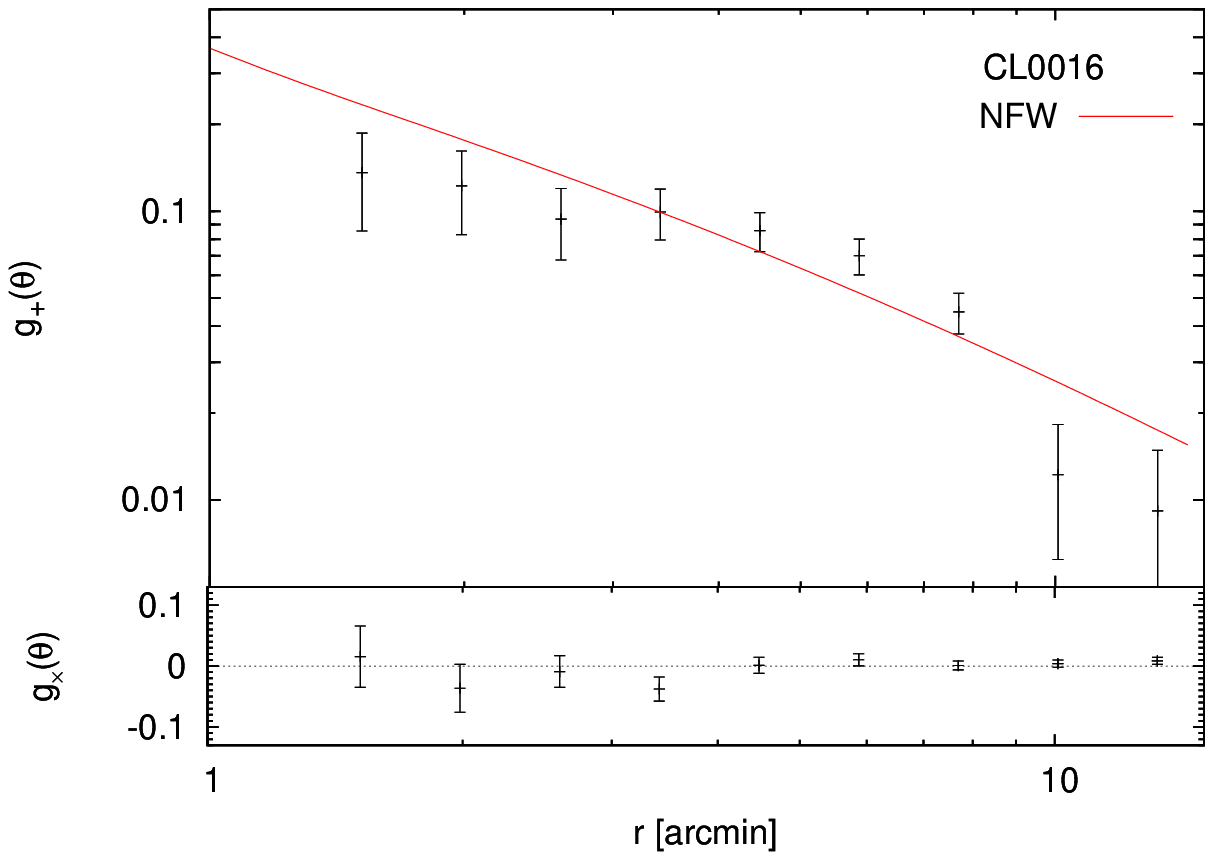}\label{fig.cl0016_nfw}}\hspace{1cm}
\subfigure{\includegraphics[width=0.8\columnwidth]{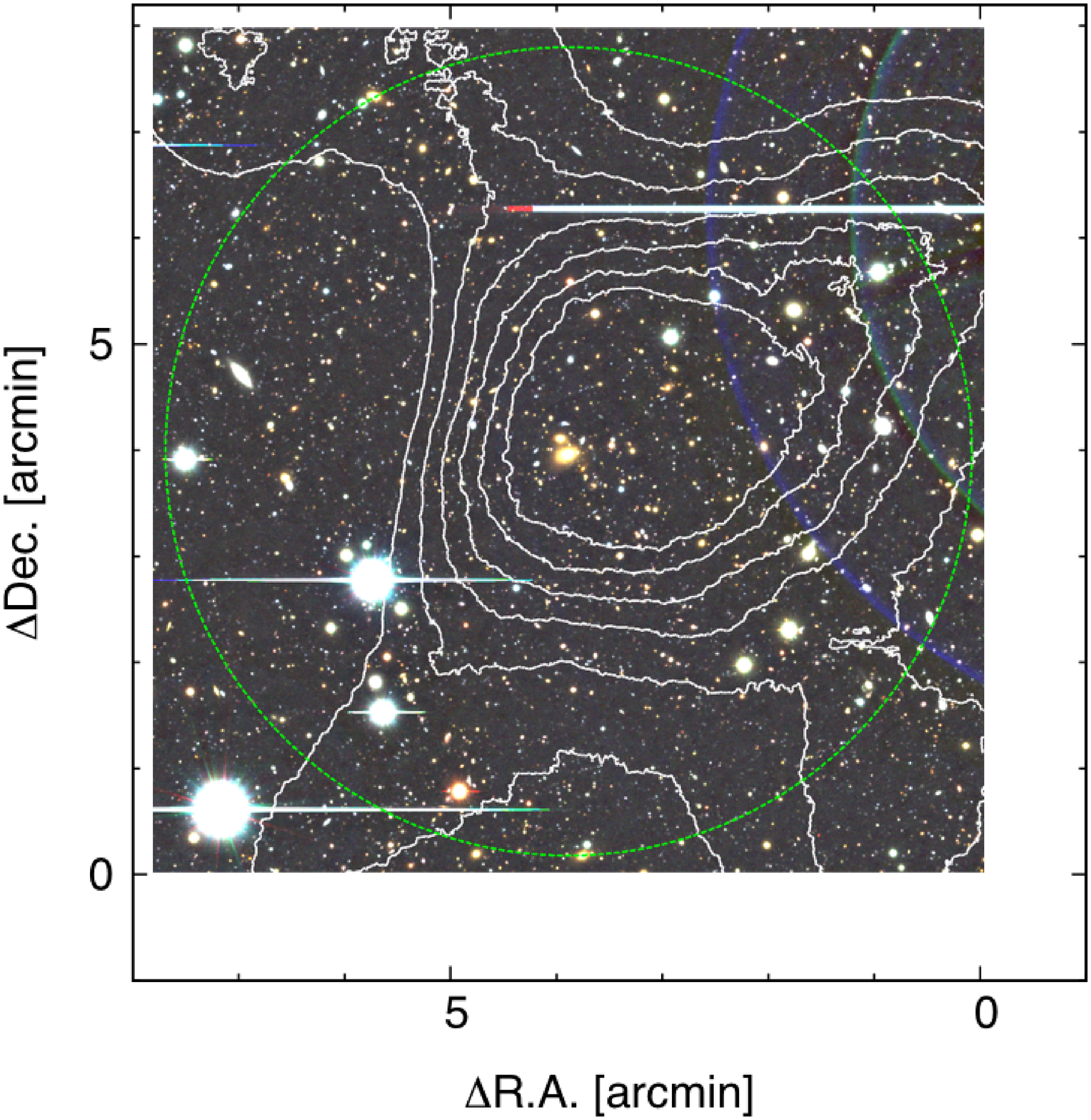}\label{fig.rxj183_contour}}\hspace{1cm}
\subfigure{\includegraphics[width=1.1\columnwidth]{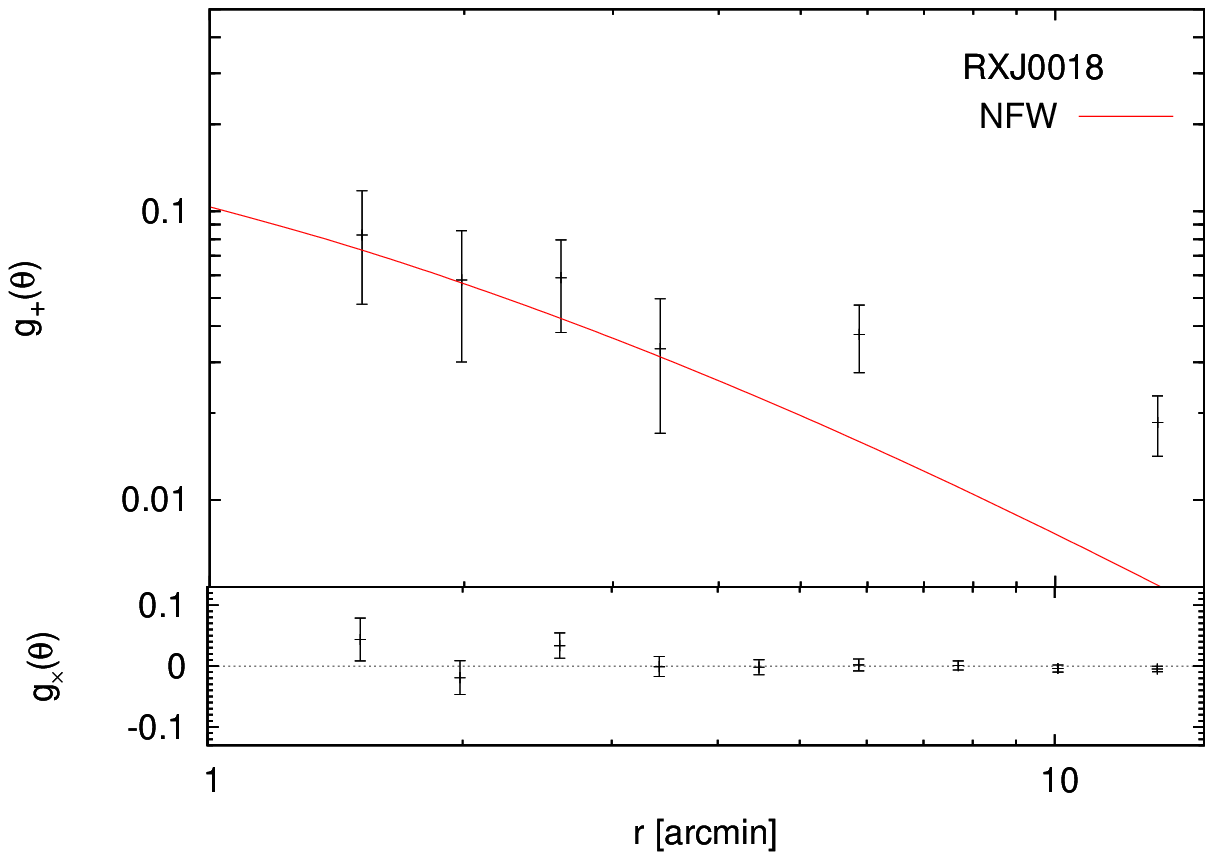}\label{fig.rxh183_nfw}}\hspace{1cm}
\caption{Left panels show the Subaru/Suprime-Cam image and contours of the convergence field for CL0016 (upper) and RX J0018 (lower). The contours (solid line) show the significance above the mean of the field edge spaced in a unit of $0.5\sigma$ from $0.5\sigma$ up to $3\sigma$ for clarity. The dashed circles indicate the viral radii estimated from the shear profile fitting in Section~\ref{sec.ana.obs}. Right panels show the radial profiles of tangential and cross shear as a function of distance from the each BCG position. The error bars show the $1\sigma$ measurement errors. The solid lines show the best-fit NFW profiles. In the tangential shear profile for RX J0018, the values of the fifth, seventh and eighth bins are not displayed due to low values.}
\label{fig.nfw_fit}
\end{center}
\end{figure*}

First, we select the Brightest Cluster Galaxy for each cluster (hereafter BCG) and define a line connecting both BCG positions, which we call BCG-BCG axis. 
%The shear profiles of the clusters are calculated with respect to the BCG position and fit with the NFW profile. 
Then, we define a filament region, which are the region between both BCGs and outside a radius R$_i$ from each BCG (shaded region in Figure~\ref{fig.exreg}). The filament length $l$ is defined on the sky plane as
\begin{equation}
L_{\rm 2D}=\sum_i{\rm R}_i+l,
\end{equation}
where $L_{\rm 2D}$ is the distance between both BCG positions and the index labels the clusters. As explained in Section~\ref{sec.filpro}, shear profiles of the filament region is calculated as a function of distance from the BCG-BCG axis.

Since lensing signals from filaments are weak, clusters at both edges of the filament can affect lensing signals of the filament significantly. Therefore, it is important to subtract such cluster contributions. In order to subtract the effect, we estimate a background which comes from cluster weak lensing signals, and subtract the background from the lensing signals in the filament region. The background is estimated with two different methods to check robustness against background subtraction methods. The first method uses analytical profiles of the clusters. The second method estimates the background from the opposite sides of the filaments. Figure~\ref{fig.exreg} illustrate the procedure which we explained in this subsection.

\subsubsection{Subtracting the analytical profiles}
The first background subtraction method is to subtract cluster lensing signals estimated from the best-fit analytical model. Assuming that the lensing signals of the clusters are well modeled with the spherically symmetric NFW profile, the effects from both clusters to the lensing signals of the filament region can be removed by subtracting the lensing signals constructed with the best-fit NFW models. 

We fit tangential shear profiles of both clusters with the NFW profile by minimizing chi-square values defined in equation~(\ref{eq.chisquare}). Figure~\ref{fig.nfw_fit} shows the tangential shear profiles and the best-fit NFW profiles. The best-fit parameters are ${\rm M_{vir}}=\left(2.09^{+0.10}_{-0.21}\right)\times10^{15}h^{-1}{\rm M}_\odot$ and ${\rm c_{vir}}=5.56^{+0.63}_{-0.79}$ for CL0016, and ${\rm M_{vir}}=\left(4.51^{+1.13}_{-1.05}\right)\times10^{14}h^{-1}{\rm M}_\odot$ and ${\rm c_{vir}}=4.61^{+1.90}_{-1.35}$ for RX J0018. The best-fit masses are consistent with the results obtained from the X-ray observations \citep{1995ApJ...448L..93H, 2007A&A...476...63S}. The two-dimensional lensing shear maps of the clusters are constructed with these parameters. Then, for each background galaxies inside $3{\rm R_{vir}}$ from BCG positions shear values of the cluster components are subtracted to remove the cluster contribution. 
We also estimate the cluster masses only with galaxies which do not exist in the filament region. The estimated parameters are almost consistent within $1\sigma$. 

\subsubsection{Empirical background subtraction using opposite sides of the clusters}
The second background estimation method is to subtract the the shear profile on the opposite sides of the clusters.
An advantage of this second approach is that it takes account of elongations of clusters which are ignored in the first method. 
As shown in Figure~\ref{fig.exreg} (stripe region), the background is set to the same length as that of the filament length. The regions inside ${\rm R}_i$ from each BCG are excluded. Then, the tangential shear in the background is estimated as a function of distance from the line along the BCG-BCG axis, which is subtracted from the tangential shear profile in the filament region to obtain the final profile.

%%%%%%%%%%%%%%%%%%%%%%%result %%%%%%%%%%%%%%%%%%%%%%%%%%%%%%%%%
\section{Result}
\label{sec.result}
\begin{figure}
\begin{center}
\includegraphics[width=8cm]{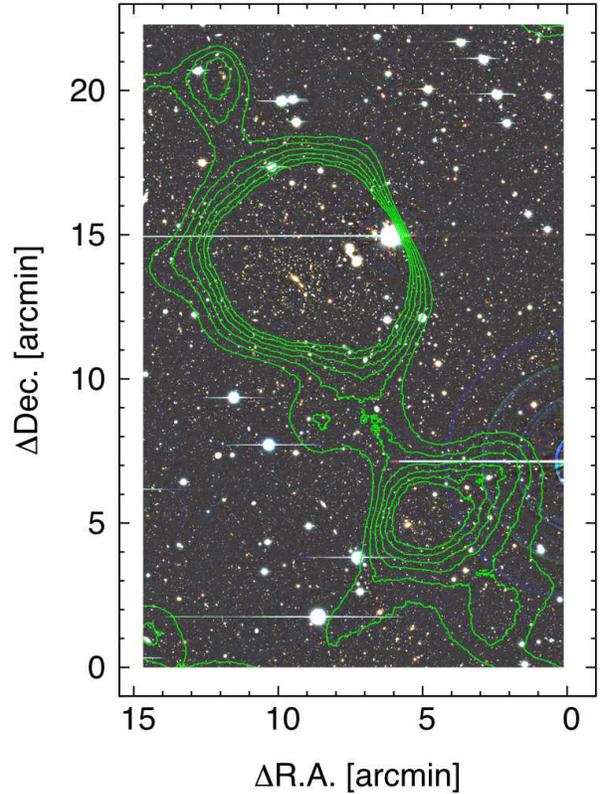}
\caption{The mass map reconstructed with weak lensing. The cluster near the centre is CL0016. RX J0018 is located on the lower-right corner. The contours show significance above the mean of the field edge spaced in a unit of $0.5\sigma$ from $0.5\sigma$.}
\label{fig.con-ima}
\end{center}
\end{figure}

Figure~\ref{fig.con-ima} shows the convergence map. We find the mass bridge between CL0016 and RX J0018, which is also seen in the galaxy distribution (Figure~\ref{fig.gal_dis}).

\begin{figure*}
\begin{center}
\subfigure{\includegraphics[width=0.9\columnwidth]{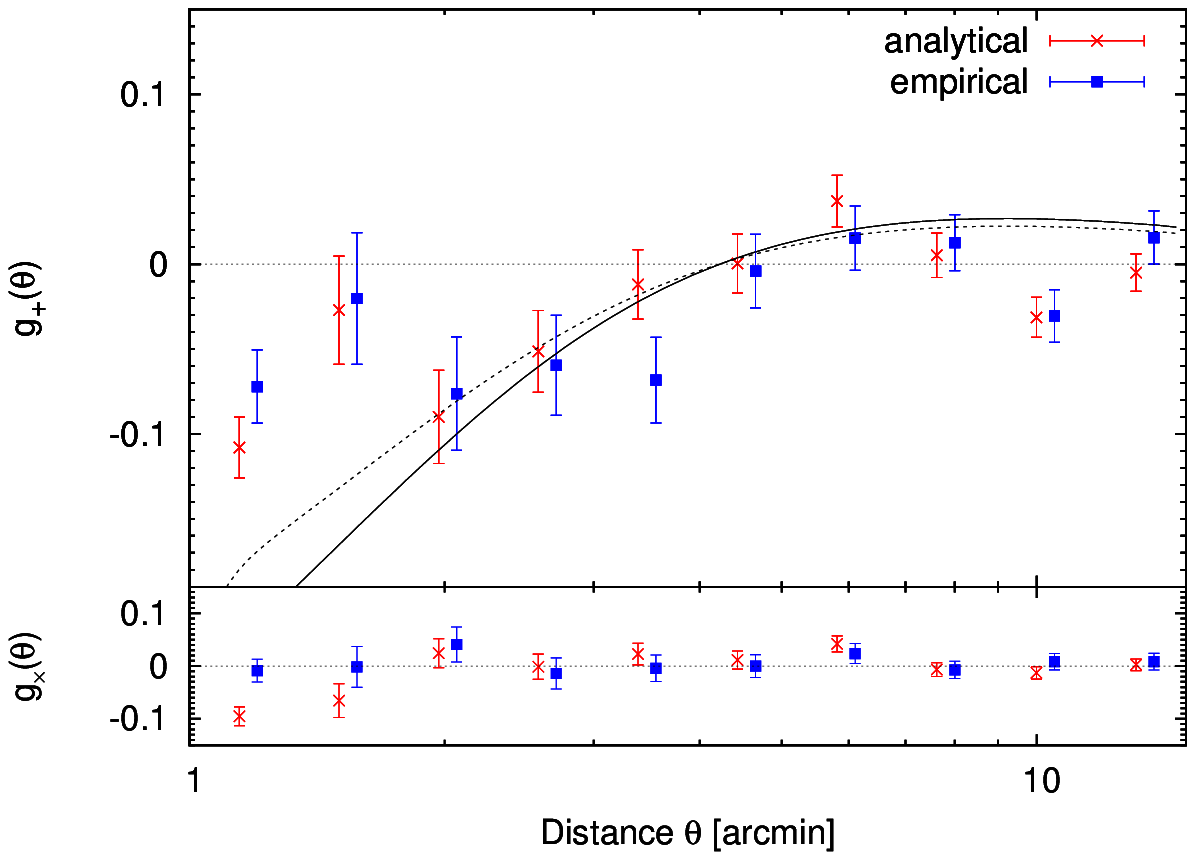}}\hspace{1cm}
\subfigure{\includegraphics[width=0.9\columnwidth]{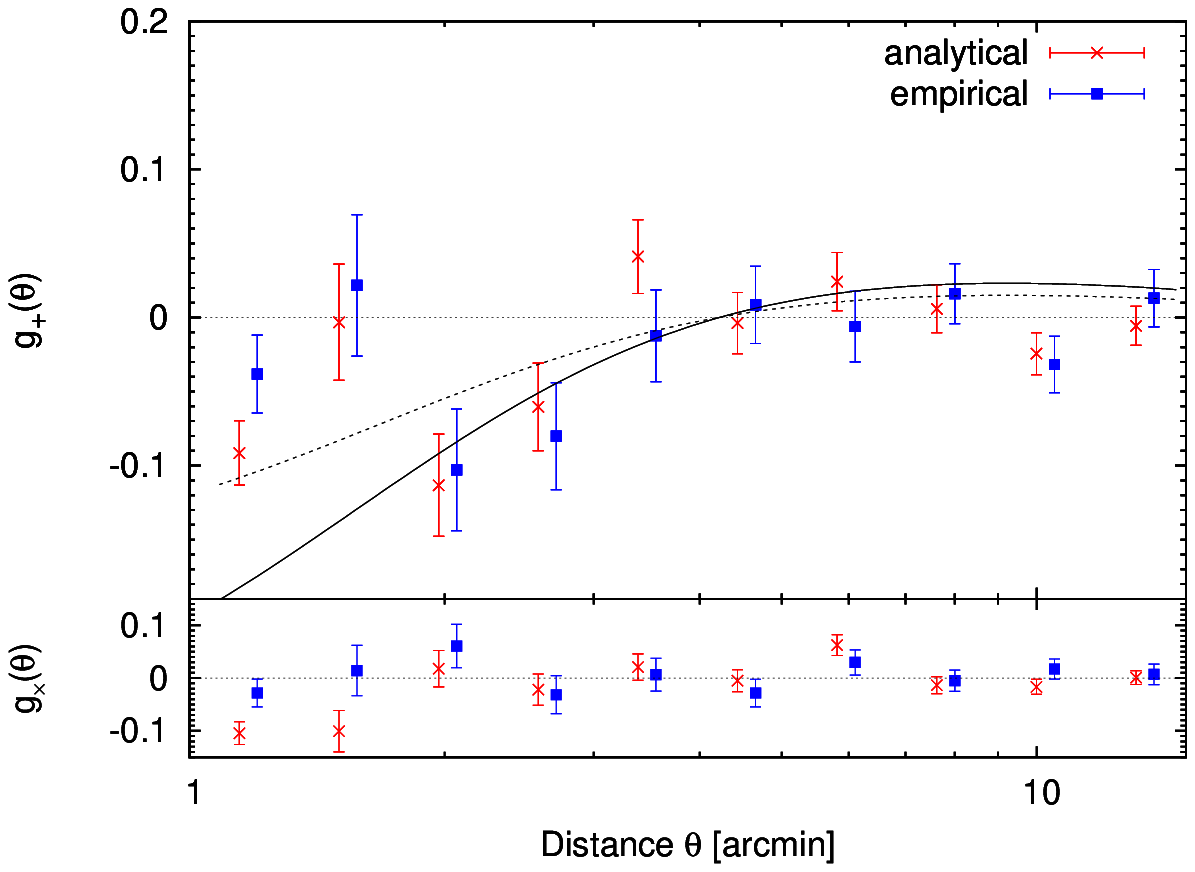}}\hspace{1cm}
\subfigure{\includegraphics[width=0.9\columnwidth]{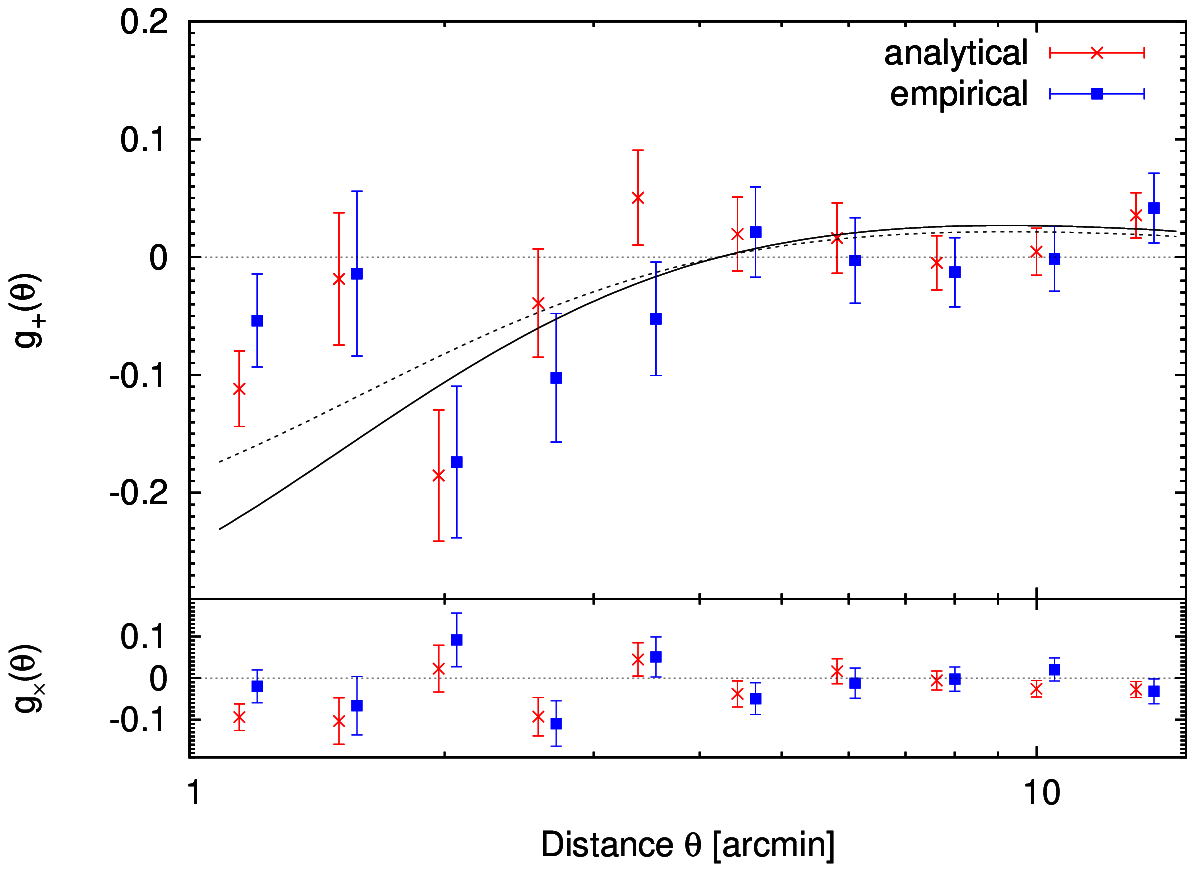}}\hspace{1cm}
%\subfigure{\includegraphics[width=0.9\columnwidth]{bgsub_bin=10_rvir_BCG_subpro2.eps}}\hspace{1cm}
\caption{Tangential profiles of the filament for different excluded regions ({\it upper left}: $1/4{\rm r_{vir}}$; {\it upper right}: $1/2{\rm r_{vir}}$; {\it lower}: $3/4{\rm r_{vir}}$). The crosses show the values estimated with the analytical background subtraction method. The squares show the values for the empirical background subtraction method. The error bars show 1$\sigma$. The solid and dotted lines show the best-fit profiles with $\theta_c=2$ arcmin for the analytical and empirical background subtracted profiles. }
\label{fig.ana_filpro}
\end{center}
\end{figure*}

Figure~\ref{fig.ana_filpro} shows the tangential and cross shear profiles of the filament for both background subtraction methods. The shear profiles are shown for different definitions of  the excluded region (${\rm R}=1/4{\rm r_{vir}}$, $1/2{\rm r_{vir}}$ and $3/4{\rm r_{vir}}$) to check the sensitivity of our results on the size of the excluded region. We find that the tangential shear profiles are similar to the shear profile expected from the convergence profile in equation~(\ref{eq.filamentmodel}) (see also \citealt{2014MNRAS.441..745H}), whereas B-mode profiles are consistent with zero for both background subtraction methods. Table~\ref{tab.sn} summarizes the values of the significances for the tangential shear profiles up to $15$ arcmin for different definitions of the excluded region, estimated from $\chi^2$ for the null shear profile. We find that the filament profiles are detected at the significance $\ga2\sigma$ for the case of empirical background subtraction method. For the analytical background subtraction method, the detection significances are much higher, but B-mode profiles in the inner region slightly deviates from zero, which may indicate insufficient background subtraction due to the aspherical profiles of the clusters. To illustrate this point, we compute the detection significances also for the B-mode (cross) shear profiles, which are summarized in Table~\ref{tab.siggc}. While the $\chi^2$ values are consistent with those expected from the $\chi^2$ distribution for the empirical background subtraction method, $\chi^2$ values are significantly larger than expected for the analytical background subtraction method, suggesting significant residuals for the analytical background subtraction.

We also check the dependence of the profiles on the size of bins and the selection of galaxies. For instance, we change the size of bins and change the photometric cut to $z_{\rm photo}-\sigma\geq0.7$ or $z_{\rm photo}-\sigma\geq0.8$. However, these changes do not strongly affect our result.

\begin{table*}
\begin{center}
\caption{Significance for the tangential shear profiles up to 15 arcmin for different definitions of the excluded region. Column (1): background estimation method; Column (2)-(4): value of significance.}
\begin{tabular}{ccccc}
\hline
background subtraction method&&&size of the excluded region&\\
&&$1/4{\rm r_{vir}}$&$1/2{\rm r_{vir}}$&$3/4{\rm r_{vir}}$\\ \hline\hline
analytical model&&13.5$\sigma$&7.36$\sigma$&4.91$\sigma$\\
empirical&&5.95$\sigma$&2.05$\sigma$&1.75$\sigma$\\
\hline
\label{tab.sn}
\end{tabular}
\end{center}
\end{table*}

\begin{table*}
\begin{center}
\caption{Similar to Table~\ref{tab.sn}. But for the cross shear profiles. Column (1): background estimation method; Column (2)-(4): value of significance.}
\begin{tabular}{ccccc}
\hline
background subtraction method&&&size of the excluded region&\\
&&$1/4{\rm r_{vir}}$&$1/2{\rm r_{vir}}$&$3/4{\rm r_{vir}}$\\ \hline\hline
analytical model&&8.29$\sigma$&8.10$\sigma$&3.30$\sigma$\\
empirical&&1.11$\sigma$&0.25$\sigma$&0.65$\sigma$\\
\hline
\label{tab.siggc}
\end{tabular}
\end{center}
\end{table*}

\begin{table*}
\begin{center}
\caption{Average projected mass density of the filament. The mass density is estimated with convergence values within 3 arcmin from the BCG-BCG line for each definition of the excluded region. The errors indicate $1\sigma$ error.}
\begin{tabular}{c|ccc}
\hline
Definition of the excluded region&$1/4{\rm r_{vir}}$&$1/2{\rm r_{vir}}$&$3/4{\rm r_{vir}}$\\ \hline\hline
Projected mass density [$10^{14}h$M$_\odot$/Mpc$^2$]&$3.23\pm0.10$&$3.21\pm0.10$&$3.20\pm0.10$\\
\hline
\label{tab.mass}
\end{tabular}
\end{center}
\end{table*}

\begin{table*}
\begin{center}
\caption{Best-fit parameters with $\theta_c=2$ arcmin. Column (1): background estimation method; Column (2)-(4): value of $\kappa_0$.}
\begin{tabular}{ccccc}
\hline
background subtraction method&&&size of the excluded region&\\
&&$1/4{\rm r_{vir}}$&$1/2{\rm r_{vir}}$&$3/4{\rm r_{vir}}$\\ \hline\hline
analytical model&&0.33&0.29&0.33\\
empirical&&0.28&0.19&0.27\\
\hline
\label{tab.parameters}
\end{tabular}
\end{center}
\end{table*}

\begin{table*}
\begin{center}
\caption{Projected mass density of the filament estimated by equation~(\ref{eq.filamentmodel}) with the best-fit parameter. The masses within $3$ arcmin from the BCG-BCG axis are calculated with $\theta_c=2$ arcmin. The error indicates $1\sigma$ error. Column (1): background estimation method; Column (2)-(4): projected mass for each excluded region definition.}
\begin{tabular}{ccccc}
\hline
background subtraction method&&&projected mass density [$10^{14}h$M$_\odot$/Mpc$^2$]&\\
&&$1/4{\rm r_{vir}}$&$1/2{\rm r_{vir}}$&$3/4{\rm r_{vir}}$\\ \hline\hline
analytical model&&$6.82^{+0.09}_{-0.09}$&$5.89^{+0.11}_{-0.12}$&$6.82^{+0.16}_{-0.16}$\\
empirical&&$5.68^{+0.12}_{-0.13}$&$3.82^{+0.15}_{-0.17}$&$5.47^{+0.20}_{-0.21}$\\
\hline
\label{tab.massana}
\end{tabular}
\end{center}
\end{table*}

We also estimate the average projected mass of the filament from the convergence map, using convergence values up to $\theta_0=3$ arcmin from the BCG-BCG axis. Here we use the convergence map before the background subtraction.
Table~\ref{tab.mass} shows the average and error of projected mass density for different excluded regions. We find that the value of the projected mass density is almost consistent with previous weak lensing studies of filaments \citep{2012Natur.487..202D, 2012MNRAS.426.3369J}. We fit the tangential shear profiles with the analytical profile defined with equation~(\ref{eq.filamentmodel}). For simplicity, we fix $\theta_c=2$ arcmin. Table~\ref{tab.parameters} shows the best-fit values for $\kappa_0$. Table~\ref{tab.massana} shows the projected mass within $3$ arcmin from the BCG-BCG axis, which are estimated by equation~(\ref{eq.filamentmodel}) with the best-fit parameters. The estimated mass densities are higher than those estimated from the convergence values that are estimated from the smoothed mass map. The fitting with the fixed parameter $\theta_c$ might also be a part of reasons for this discrepancy. 

%%%%%%%%%%%%%%%%%%%%%%%%conclusion%%%%%%%%%%%%%%%%%%%%%%%%%%%%%%%%%%
\section{Conclusion}
\label{sec.conclusion}
In this paper, we have investigated the weak lensing properties of a filament between the cluster CL0015.9+1609 and RX J0018.3+1618 with the Subaru/Suprime-Cam data. We have selected background galaxies with photometric redshift and estimated the lensing signals with $Lensfit$. We have subtracted the effects of the clusters from the lensing profile of the filament with two different methods. One is to subtract the best-fit NFW profiles, and the other is to subtract the profile obtained on the opposite sides of the filament.

We have fitted the tangential shear profiles of the clusters with the NFW model, and obtained ${\rm M_{vir}}=\left(2.09^{+0.10}_{-0.21}\right)\times10^{15}h^{-1}{\rm M}_\odot$ and ${\rm c_{vir}}=5.56^{+0.63}_{-0.79}$ for CL0016, and ${\rm M_{vir}}=\left(4.51^{+1.13}_{-1.05}\right)\times10^{14}h^{-1}{\rm M}_\odot$ and ${\rm c_{vir}}=4.61^{+1.90}_{-1.35}$ for RX J0018. The estimated viral radii are used to define the size of the excluded regions.

We have found the mass bridge between two clusters on the convergence map. We have estimated the average surface mass density for three different excluded region definitions. The surface mass density for the excluded region definition of ${\rm R=3/4r_{vir}}$ is found to $(3.20\pm0.10)\times10^{14}h$M$_\odot$Mpc$^{-2}$ from the convergence map. This is consistent with previous studies of filaments.

We have derived the shear profiles of the filament using different excluded region definitions and background subtraction methods. For all the definitions of the excluded region, we have found that the cross shear values are almost consistent with zero and tangential shear profiles have same profiles as those expected from simulations, indicating that the lensing signals of the filament are detected. We have compared the shear profiles with a null profile up to $15$ arcmin from the BCG-BCG axis to find that the lensing signals of the filament are detected at the significance $\ga2\sigma$ in all cases.

The detection of filaments with weak gravitational lensing is difficult due to small values of lensing signals. However, large-scale surveys such as Hyper Suprime-Cam and Large Synoptic Survey Telescope may enable direct observations of more filaments. Statistics of weak lensing properties of filaments therefore serve as a new test of structure formation in the Universe \citep{2014MNRAS.441..745H}.
%%%%%%%%%%%%%%%%%%%%%%%acknowledgement
\section*{Acknowledgements}
%We thank an anonymous referee for careful reading and suggestions to improve the quality of the article.
We would like to thank Takashi Hamana, Satoshi Miyazaki, Yuki Okura and Junko Ueda for useful discussion. We would like to thank Tadayuki Kodama for providing the photo-z catalogue. Data analysis were carried out on common use data analysis computer system at the Astronomy Data Center, ADC, of the National Astronomical Observatory of Japan. Based on data collected at Subaru Telescope and obtained from the SMOKA, which is operated by the Astronomy Data Center, National Astronomical Observatory of Japan. This work was supported in part by the FIRST program Subaru Measurements of Images and Redshifts (SuMIRe), World Premier International Research Center Initiative (WPI Initiative), MEXT, Japan, and Grant-in-Aid for Scientific Research from the JSPS (26800093), and in part by Grant-in-Aid for Scientific Research from the JSPS Promotion of Science (23540324, 23740161).

%%%%%%%%%%%%%%%%%%%%% reference %%%%%%%%%%%%%%%%%%%%%%%%%%%%%
\bibliographystyle{mn2e}
\bibliography{mn-jour,bibtex}
\end{document}